\documentclass[ ]{copernicus2}

\usepackage{color}
\usepackage[T1]{fontenc}

\begin{document}

\title{Conductivity spectrum and dispersion relation\\ in solar wind turbulence%by seed-field generation
}

\author[1]{R. A. Treumann\thanks{Visiting the International Space Science Institute, Bern, Switzerland}}
%\author[3]{R. Nakamura}
\author[2]{W. Baumjohann}
%\author[2]{Y. Narita}

\affil[1]{Department of Geophysics and Environmental Sciences, Munich University, Munich, Germany}
%\affil[2]{Department of Physics and Astronomy, Dartmouth College, Hanover NH 03755, USA}
\affil[2]{Space Research Institute, Austrian Academy of Sciences, Graz, Austria}
%% The [] brackets identify the author to the corresponding affiliation, 1, 2, 3, etc. should be inserted.

\runningtitle{Magnetic turbulence in the solar wind}

\runningauthor{R. A. Treumann and W. Baumjohann}

\correspondence{R. A.Treumann\\ (art@geophysik.uni-muenchen.de)}

\received{ }
%\pubdiscuss{ } %% only important for two-stage journals
\revised{ }
\accepted{ }
\published{ }

%% These dates will be inserted by the Publication Production Office during the typesetting process.

\firstpage{1}

\maketitle

{\bf{Abstract}}. -- Magnetic turbulence in the solar wind is treated from the point of view of electrodynamics. This can be done based on the use of Poynting's theorem attributing all turbulent dynamics to the spectrum of turbulent conductivity. For two directions of propagation of the turbulent fluctuations of the electromagnetic field with respect to the mean plus external magnetic fields an expression is constructed for the spectrum of turbulent dissipation. Use of solar wind observations of electromagnetic power spectral densities in the inertial subrange then allows determination of the conductivity spectrum, the dissipative response function, in this range. It requires observations of the complete electromagnetic spectral energy densities including electric power spectral densities. The dissipative response function and dispersion relation of solar wind inertial range magnetic turbulence are  obtained. The dispersion relation indicates the spatial scale decay with increasing frequency providing independent support for the use of Taylor's hypothesis. The dissipation function indicates an approximate shot-noise spectrum of turbulent resistivity in the inertial range suggesting progressive structure formation in the inertial range which hints on the presence of discrete mode turbulence and nonlinear resonances.

\vspace{0.5cm}
\section{Introduction}
Magnetic turbulence has, in the near-past, made substantial progress both in theory and observation. This became possible mainly by the availability of sufficiently sophisticated instrumentation aboard spacecraft in the solar wind as well as the increase in computer power. The solar wind is though not the ideal so nevertheless a very good place to study magnetic turbulence in situ. It streams at supersonic/super-alfv\'enic speed out from the solar corona into the heliosphere, is fairly stationary, transports any disturbances at high speed downstream. Any spacecraft located at some point approximately at rest with respect to the solar wind is thus swamped with turbulence which crosses over its location and can monitor its spectrum. This has now been done quite frequently. For an early recollection which already contained the main results at that time the reader my consult \citep{gold1995}. Fluctuation spectra are usually measured at one single spacecraft, mostly yielding because of the simplicity of measurement magnetic spectral power densities. These are obtained from the observed magnetic fluctuations in a certain frequency range. Under the so-called Taylor hypothesis \citep{taylor1938} that the turbulence is frozen to the flow such spectra can straightforwardly be transformed into (one-dimensional) wavenumber spectra, if only the speed of the flow substantially exceeds any of the internal turbulent group velocities. In high speed solar wind this assumption is valid for a large range of frequencies respectively turbulent wave numbers. Of course, the Taylor hypothesis does not apply into a direction substantially oblique to the flow. The main result of these observations and the application of the Taylor hypothesis are spectra of the magnetic power as function of either frequency $\omega$ (covering the complete turbulent magnetic field in the respective frequency range) or one-dimensional wavenumber $\mathbf{k}$. Such power spectra yield the wanted spectral slope of the power spectral density which is constant over the extension of the inertial range frequency/wavenumber interval as predicted by Kolmogorov \citep{kol1941} for fluid turbulence about one century ago. Kolmogorov's global theory was extended twenty years later to include magnetohydrodynamic turbulence \citep{irosh1963,krai1964}. A recollection is found in \citep{bis2003}. 

Identification of power spectral slopes meanwhile occupies a vast literature. Observations in the solar wind are claimed to provide robust spectral slopes (spectral indices) $\gamma\approx 5/3$ suggesting that Kolmogorov-like inertial sub-ranges are indeed realized in the solar wind. Their extension in frequency covers roughly one order of magnitude. Spectral breaks have also been detected as well as the transition from the inertial into the so-called dissipative range where the magnetic spectral power steeply decays, becoming dissipated by some still badly defined processes. Theory meanwhile progressed in two direction. Analytical theory went on to investigate magnetic turbulence (which in the lowest frequency range below the ion-cyclotron frequency $\omega_{ci}=eB/m_i$ is in the magnetohydrodynamic range) based on Elsasser variables $\mathbf{z}^\pm=\mathbf{v\pm B}/\sqrt{\mu_0\rho}$ which mix velocity $\mathbf{v}$ and magnetic $\mathbf{B}$ fields with $\rho=(m_i+m_e)N$ the mass density, $N$ quasi-neutral number density, and $m_{i,e}$ ion and electron mass. These coordinates account for the two directions of propagation in magnetohydrodynamics. Combined with helicity they also include polarizations. The most elaborate set of expressions for turbulent correlations including dissipation was given in \citep{zank2012}. Weak turbulence expressions based on full kinetic theory for parallel propagation only are found in \citep{yoon2007a,yoon2007b}. All these expressions are highly complex and thus very difficult to apply to observations. Numerical simulations are also ubiquitous \citep{berez2011,berez2014,bold2013,zhdan2014,zhou2004}. They provide, similarly to observations, spectra which allow for the determination of spectral slopes. They also provide pictures of turbulent flows and correlations of various quantities, in principle permitting comparison with theory.

The main question when considering observations is: What is the physical meaning of spectral slopes?  Can anything physically more substantial be deduced from their detection in view of the physics of turbulence? Simply stating that in a certain interval of frequencies (or under some conditions spatial scales) one observes an approximate inertial range in agreement with Kolmogorov (or with some other prediction) is not very satisfactory. Detection of spectral breaks is interesting, but the simple statement it means Kraichnan-Iroshnikov \citep{irosh1963,krai1964} scaling, i.e. Alfv\'enic dominance, is rather meagre. Detection of the transition to the dissipative range is most interesting and important as well, as long as the physical scale can be extracted, but what can otherwise physically be inferred from it? What do slopes tell us more than that there are slopes of a certain measured relatively frequently repeating steepness? What is the meaning of slopes in the dissipation range?

Concerning the Kolmogorov slope it is frequently argued that they are robust. Robust against what? It means they are detected all times and under very different conditions. Recently Chen et al. \citep{chen2011} based on {\small ARTEMIS} measurements in the lunar solar wind lagrangean point analysed a large number of time intervals of comparably slow-solar wind flow turbulence and determined the inertial-range slopes. Their results show that the slopes have a wide spread 
\begin{equation}
1.5\lesssim\gamma_B\lesssim1.8
\end{equation}
though their mean value is surprisingly close to Kolmogorov's canonical $\gamma_B=5/3$. This is considered to be robust. However, we know from thermodynamics, statistical mechanics, and particularly from chaos theory \citep{beck1997} that rather small deviations in spectral slopes frequently imply vast differences in the physical settings. They may imply intermittency and even imply phase transitions. 

Of course, magnetic turbulence theory is very difficult. The problem about it is not only that it mixes the dynamics of the charged matter with the dynamics of the electromagnetic field. It also is highly nonlinear with all spatial scales interconnected in a way which to large extent inhibits a perturbative treatment. Even weak low-frequency/large-scale turbulence which was recently treated \citep{yoon2007a,yoon2007b} perturbatively within kinetic theory yields as well very involved expressions hardly suited for use in experimental work. On the other hand, non-perturbative treatments are sparse though treatments based on the renormalization group theory can be found \citep{verma2004} while producing intractable expressions. In an earlier paper we attempted an inverse-scattering approach to turbulence \citep{treumann2016}.

In the following we treat magnetic turbulence from the point of view of electrodynamics. In this we follow \citep{treumann2017} by assuming that all of the dynamics is contained in the transport coefficients. We show that reference to some of the observations allows to draw conclusions about the spectrum of the plasma response to the presence of electromagnetic turbulence which permits some conclusions about the field behavior. In addition it enables us to provide an expression for the turbulent dispersion relation that describes the electromagnetic waves in the inertial range. 

\section{Turbulent response function}
Textbook physics \citep{landau1984} using Poynting's theorem of the transport of electromagnetic energy in a dissipative (electrodynamically active) medium can be used in order to reduce the complete set of electrodynamic equations to the linear form of Poynting's theorem describing  the energy density in turbulent electromagnetic fluctuations. This transport is governed by the equation \citep{treumann2017}
 \begin{equation}\label{eq-one}
\frac{\partial}{\partial t}\bigg[\frac{B^2(t)}{\mu_0}-\int\limits_{-\infty}^t\mathrm{d}t'\mathbf{E}(t')\cdot\mathbf{\sigma\!\!\!\!\sigma}(t')\cdot\mathbf{E}(t')\bigg]=0
\end{equation}
where $\mathbf{B,E}$ are the magnetic and electric fluctuation fields which are functions of time $t$, and $\mathbf{\sigma\!\!\!\!\sigma}(t)$ is the turbulent electric conductivity of the plasma, which is considered to be completely unknown. It contains the total contributions of the dynamics as well as the response of the matter to the presence of the turbulent fields. We do not make any assumptions about the underlying dynamics at this place as we are interested in obtaining an expression which relates the turbulent conductivity to measurable quantities. 

The assumptions under which the above equation holds are as follows. Turbulence is considered to be low frequency, far below plasma frequency. This allows to neglect all relativistic displacement current effects. The plasma is assumed quasineutral with no electric Coulomb fields involved, which holds as long as the scales are sufficiently large and time-scales far below plasma frequency. Thus one may use the Lorentz radiation gauge. This is justified as long as magnetic turbulence is considered and electrostatic fields are not involved, the case applicable for instance to solar wind turbulence and excluding kinetic theory which comes into play at microscopic scales. Moreover, it is assumed that all mean fields that result from averaging over the fluctuation scales in time and space are constant on the fluctuation scales. In this case a mean stationary magnetic field is allowed to be present and may also contain a stationary external field. 

This equation will be treated in two cases, where its structure is retained. The first case is that all electromagnetic fluctuations propagate solely parallel to the ambient mean magnetic field. Since magnetic turbulence under the above assumptions is solenoidal, i.e. completely electrodynamic such that for the fluctuations $\mathbf{E\times B}=0$, parallel propagation implies that $\mathbf{E,B}$ are both perpendicular to the mean ambient field $\mathbf{B}_0$ while the wavenumbers are all $\mathbf{k}||\mathbf{B}_0$. In this case one has
\begin{equation}
\mathbf{E\cdot\sigma\!\!\!\!\sigma\cdot E}= \sigma^TE^2
\end{equation}
where $\sigma^T$ is the transverse turbulent conductivity which contains the transverse response. In the second case we assume perpendicular propagation $\mathbf{k}\perp\mathbf{B}_0$ and, in addition $\mathbf{E}||\mathbf{B}_0$. Then we have
\begin{equation}
\mathbf{E\cdot\sigma\!\!\!\!\sigma\cdot E}= \sigma^\|E^2
\end{equation}
We may understand the fields in both cases to be averaged about the azimuthal angle around the mean field. The other transverse case $\mathbf{B}||\mathbf{B}_0$ we do not consider because it is more difficult to treat as no such assumptions can be made. Thus our theory is incomplete also to this extent as it excludes the magnetically compressive component which, however, can easily be distinguished. 

The two above cases can be treated similarly. We do this for the transverse case. Fourier transforming Eq. (\ref{eq-one}) with respect to time yields
\begin{equation}
(B^2_\perp)_\omega=\frac{i\mu_0}{\omega}(E_\perp^2\sigma^T)_\omega
\end{equation}
where the index $\omega$ indicates that the quantity is the Fourier transform and depends on frequency $\omega$.
Assume that the turbulence is stationary. Then all quantities depend only on time differences $t-t'$. The above time integral becomes a convolution integral, and the temporal Fourier transform reduces to the product of Fourier transforms. We do, however, not apply this rule to the fields but only to the product of fields and turbulent conductivity. We are not interested in the fields themselves. These can be measured, while the conductivity is unknown. Implictly we thereby assume that both spectra, the magnetic and electric, have been measured. Then we find for the transverse as well as for the parallel case
\begin{equation}\label{eq-sigma}
\sigma^{T,\|}_\omega=-\frac{i\omega}{\mu_0}\frac{(B_\perp^2)_\omega}{(E^2_{\perp,\|})_\omega}
\end{equation}

These expressions hold for the two cases of different propagation and could be used independently. Of course, they can be distinguished if the observations differentiate between parallel and perpendicular electric field fluctuations. The case of transverse electric field fluctuations can be separated as well by excluding mean-field-aligned (compressive) magnetic components. Hence, once this can all be done one is in the position to determine the above transverse and parallel frequency spectra of turbulent conductivities by taking the ratios of the corresponding power spectra of the magnetic and electric field components at each frequency. Such a procedure provides new information about the state of the turbulence. It might be used to infer more about the underlying tubulent dynamics. 

Theoretically it is simple matter to obtain the time dependence of the turbulent conductivity by Fourier-retransformation from the frequency into the time domain. In doing this one should keep in mind that only time differences $\tau=t-t'$ make sense, because stationarity of the turbulence has been assumed. Therefore, dealing with the turbulent conductivity spectrum in frequency space instead of time provides more physical information.

\section{Application to observed turbulent power laws}
In order to determine the turbulent conductivity spectrum, observations of both the electric and magnetic fields are required. To our knowledge such observations have only rarely been performed in the past. There are two published cases \citep{bale2005,chen2011} to which we may refer. Unfortunately, neither of these observations distinguish between the directions of the electric and magnetic fluctuation fields with respect to the mean and ambient magnetic fields. Moreover, all observations are single spacecraft measurements, and what concerns the electric power spectra, the Cluster measurements of \citep{bale2005} refer only to the spectrum obtained in the spacecraft frame. However, the electric field is not covariant, i.e. it depends on the frame in which it is measured. Electric field fluctuations $\mathbf{E}'$, in order to become comparable to the frame-independent (covariant) magnetic fluctuations $\mathbf{B}$ should be transformed into the (unprimed) streaming frame, in this case the solar wind, by applying the Lorentz transformation
\begin{equation}
\mathbf{E}=\mathbf{E}'+\mathbf{v}_{sw}\times\mathbf{B}
\end{equation}
The latter term vanishes only if all the magnetic fluctuations are parallel to the solar wind velocity and its fluctuations. In a high-velocity stream the velocity in the latter term may be approximated by the mean streaming speed. If the flow is slower than the velocity fluctuations must be included. This implies measuring the ion distribution function and determining the density and velocity thus putting high standards of measurement on the particle instruments and implying susceptible uncertainties which can be reduced by statistically considering many events. 

This has been done for the {\small ARTEMIS} measurements of the fluctuating turbulent electric fields in the solar wind by Chen et al. \citep{chen2011}. These authors find a roughly one decade wide ``inertial subrange''  in their measurements for a large number of time slots of observations in the solar wind. Both the magnetic and electric powers in this range obey power laws $\sim\omega^{-\gamma}$. The mean inertial magnetic power law in this range is about Kolmogorov, $\gamma_B\approx5/3$, while the mean inertial electric power law in the proper frame of the solar wind flow is non-Kolmogorov. It is found to obey $\gamma_E\approx3/2$. These values though considered to be robust must still be taken with some care for all the above mentioned reasons, in particular the large scatter of the individual spectral slopes for each observation-time slot. 

In spite of the possible reservations based on the above arguments, we can make use of these singular observations in order to determine a proxy of the turbulent conductivity spectrum. Since the measurements do not distinguish between the components, we may use either of the expressions for transverse or parallel conductivity spectra. 

Inserting into (\ref{eq-sigma}), the turbulent conductivity spectrum in the proper frame of the solar wind in the inertial range is itself a power law. Since amplitudes play no role for our purposes (though of course could be used to infer about the Kolmogorov constant for the two fields), this power law is given by
\begin{equation}
\sigma_\omega\propto \omega^{-\gamma_B+1+\gamma_E} =\omega^{-\gamma_\sigma}
\end{equation}
in the inertial range, yielding 
\begin{equation}
\gamma_\sigma=\gamma_B-1-\gamma_E\approx -\frac{5}{6}
\end{equation}
The resulting inertial range power law turbulent conductivity spectrum $\sigma_\omega\propto\omega^{0.83}$ implies an \emph{increase} in the turbulent conductivity with increasing frequency. Given all the mentioned uncertainties, this is pretty close to $\gamma_\sigma= -1$ which indicates that the turbulent solar wind resistivity spectrum 
\begin{equation}
\eta_\omega\sim\omega^{-1}
\end{equation}
behaves approximately like a shot-noise spectrum. In view of the straightforward way it has been obtained considering the transport of electromagnetic energy in the turbulence, it tells that the inertial range exists because dissipation \emph{decreases} in this range. Decrease of dissipation with frequency (not in time, however) is not easy to understand. It implies that at higher frequencies in the inertial range proportionally less energy is dissipated than at lower frequencies. This probably reflects the action of self-organization here, the progressive production of small-scale structure.

This result applies just to the inertial range in solar wind turbulence. Measurements are also available outside this range in the magnetohydrodynamic regime and could applied there as well in order to determine the turbulent conductivity spectrum. By applying our reasoning to all parts of the turbulent electric and magnetic power spectra and by distinguishing between the components of the fields and their projections on the main field, the turbulent conductivity spectrum in the frequency domain can, in principle, be determined in its transverse and parallel components for the entire observable spectral range outside the dissipation regime. Since this is not in our reach however, we stay with this partial result here. Finite real conductivities imply dissipation, one can therefore in principle also conclude about the various dissipations which are acting in each of the different spectral ranges. Moreover, where the Taylor hypothesis is applicable, the turbulent conductivity spectrum can be transformed into the range of spatial scales.

\section{Turbulent dispersion relation}
Determination of the turbulent conductivity spectrum permits the construction of the turbulent dispersion relation. This might not be obvious at first glance. The rigorous arguments will be given in \citep{treumann2017}. In brief, the turbulent conductivity spectrum can be transformed into the spectrum of the dielectric response tensor of the plasma which, since the conductivity results from turbulence, is a turbulent response to the presence of turbulent electromagnetic fields. Its Fourier transformed version reads simply
\begin{equation}
\mathbf{\epsilon}_\omega=\frac{2i}{\omega\epsilon_0}\mathbf{\sigma}_\omega
\end{equation}
We are dealing solely with electromagnetic waves of low frequency. The dispersion relation of such waves is given as
\begin{equation}
\mathcal{N}^2=\epsilon_\omega, \quad\mathrm{with}\quad {\mathcal{N\!\!\!\!\!\!N}}=\mathbf{k}c/\omega
\end{equation}
where $\mathcal{N\!\!\!\!\!\!N}$ is the vector  index of refraction. Here we use only its scalar version because the observations do not provide more information. Inserting from (\ref{eq-sigma}) one obtains a general expression for the turbulent dispersion relation 
\begin{equation}
\mathcal{N}^2_{\|,\perp} = \frac{2}{\epsilon_0\mu_0}\frac{(B^2_\perp)_\omega}{(E^2_{\|,\perp})_\omega}
\end{equation}
Since $E/B$ is a velocity by dimension, we could have guessed this relation from the beginning. It yields the general turbulent dispersion relation
\begin{equation}
k^2_{\|,\perp}=2\omega^2(B^2_\perp)_\omega/(E^2_{\|,\perp})_\omega
\end{equation}
which is completely determined by the observed power spectral densities as functions of frequency. This dispersion relaion is by no means a result of linear theory because $\sigma_\omega$ is a nonlinear function which contains all the nonlinear turbulent dynamics. Thus, the dispersion relation is the fully nonlinear dispersion relation of electromagnetic turbulence in terms of the nonlinear conductivity spectrum which has been given before.

In order to derive it for our observational case \emph{in the solar wind inertial range only}, we use the power indices as obtained by Chen et al. \citep{chen2011}, which is a severe restriction to a very particular set of observational data! Other observations might and should provide a different result than the one we derive below, in particular if not applied to the inertial but to other spectral ranges. 

Use of the above inferred spectral powers yields the nonlinear dependence
\begin{equation}
k^2(\omega)\propto \omega^{11/6}\quad\mathrm{or}\quad k\propto \pm\;\omega^{11/12} \sim \pm\;\omega^{0.92}
\end{equation}
which can also be written in conventional form in terms of the frequency of the turbulent fluctuations
\begin{equation}
\omega(k) \propto \pm\ k^{12/11} \sim \pm\ k^{1.091}
\end{equation}
Turbulent wave numbers in the inertial range increase roughly proportional to the frequency, obeying an approximately though not exactly linear dispersion relation. The turbulent fluctuations in the inertial range behave roughly, though again not exactly, like ordinary sound.

That scales in the inertial range decrease with increasing frequency is reasonable and agrees with any expected behavior of turbulence. In the inertial range this increase is approximately inversely proportional to the frequency, being only slightly weaker. This might not be true for other parts of the spectrum, however. It applies solely to the observed Kolmogorov-like inertial subrange in solar wind turbulence of the magnetic field, in addition being subject to the above mentioned uncertainties and restrictions.

{The above dispersion relation is based on the mean spectral indices of the magnetic and electric power spectral densities. These indices have a substantial scatter \citep{chen2011}. In order to get an impression of the effect of the scatter on the dispersion relation we use the values given for the magnetic power spectral densities while keeping the mean spectral index $\gamma_E=\frac{3}{2}$ for the electric power. This procedure yields}
{\begin{equation}
\omega(k)\propto \bigg\{
\begin{array}{ccc}
 k^4&   & (\gamma_B\sim1.5)  \\
  k^\frac{3}{2}&  &  (\gamma_B\sim1.8)
\end{array} 
\end{equation}
The mean dispersion relation is not contained in this range because we only used the fixed mean slope of the electric power spectral density.}

{The scatter of the inertial range spectral index produces vastly different dispersion relations. Of course the two limits for the spectral index are extreme values, while the means are more reliable. Moreover have we not used the scatter in the electric slopes. Nevertheless this exercise shows that the turbulent physics behind depends very sensitively on the precision of determination of the spectral slopes.}

From the above \emph{mean} dispersion relation we find the inertial range phase and group velocites of turbulent fluctuations in the solar wind. In this particular case, the solar wind inertial range, both the phase and group velocities of turbulent fluctuations obey exactly the same scaling with wavenumber:
\begin{equation}
v_\mathit{ph}=\frac{\omega}{k}\propto \pm\ k^{1/11}, \quad v_\mathit{g}=\frac{\partial\omega}{\partial k}\propto \pm\ k^{1/11}
\end{equation}
Thus, up to a numerical factor, the inertial range phase and group velocities of the turbulent fluctuations show the same dependence on spatial scale, with the group velocity being very slightly larger than phase speed by the factor $12/11\approx 1.1$. This implies run-away of shorter modes. One made note that the same behavior holds also for the two extreme above dispersion relations.

Though this is interesting to know, its physical meaning is obscured by the lack of knowledge about the turbulent dynamics. The inferred dispersion relation might, however, be subject to Kolmogorov-Zakharov turbulence \citep{zakh1992}, which can be described by a wave-kinetic equation approach, or to its extension to discrete wave turbulence and nonlinear resonances \citep{kart2010,kart2009}, which include the coupling to non-resonant modes (so-called quasi-resonances). What it suggests, however, is that the phase and group velocities increase with wavenumber and maximize towards the end of the inertial range when the spectrum enters into the dissipation range. The spectral energy dumped into the shortest inertial scales propagates at the largest group velocity. Some inferences about the dissipation range conductivity spectrum will be discussed elsewhere \citep{treumann2017}.

\section{Conclusions}
The present note dealt with a subfield in stationary magnetic turbulence. It  just considered its electromagnetic part assuming that all dynamics is covered by the conductivity which naturally is given by some kind of Ohm's law, discussion of which will be given elsewhere \citep{treumann2017}. Its intention is to relate observations of power spectral densities observed in magnetic turbulence to the dissipative response function. For two kinds of propagation of the electromagnetic fluctuations in turbulence with respect to the mean field direction this yields expressions of the turbulent conductivity in terms of the measured magnetic and electric power spectral densities. This is a weak restriction only because observationally it should be no problem to separate the three directions of the magnetic field fluctuations by determining the mean field and taking the appropriate projections. Then the two data sets which correspond to the parallel and transverse conductivities can be treated separately by the method proposed here. Analyzing the third set requires a different approach.  

It should be pointed out that only the combination of the electric and magnetic fluctuation spectra contains the complete electromagnetic information about the components of the magnetic turbulence. The observation of magnetic and electric power spectral densities is thus crucial as both are needed in order to determine the conductivity spectrum. Though the available data did not distinguish between the two directions of propagation, this could, in our case, be done using spacecraft observations in the solar wind turbulent inertial subrange.  The interesting result, which still is the outcome of a mixture of these data sets, nevertheless showed that the mixed turbulent conductivity spectrum is itself a power law. 

This power law, surprisingly, indicates that the turbulent conductivity \emph{increases} with increasing frequency towards the high-frequency end of the inertial range. This spectrum corresponds to an approximate shot-noise spectrum in the turbulent resistivity, which  indicates the formation of structure or self-organization. Obviously the inertial range is a spectral region where dissipation generates new structure. It is not precisely known what kind of structure this is. One may, however, speculate that the inertial range is the region of formation of ever smaller-scale turbulent current filaments the scales of which shrink towards the high-frequency end of the inertial range where the spectrum enters the dissipation range, thereby preparing the collisionless dissipation of the turbulent energy. This structure formation goes on the costs of the injected turbulent energy. The inertial range is effectively an open system. Structure formation and self-organization seem to reduce entropy here in order to maintain an about constant power law slope of the turbulent spectrum. The most efficient and violent process of dissipation in such narrow current elements is spontaneous reconnection \citep{treu2015}. This sets spontaneously on once the current filament scales match the electron inertial length at the end of the inertial range.

Moreover, use of the conductivity spectrum permits construction of the turbulent dispersion relation. This was done only for the available inertial range data in the solar wind. In agreement with expectation, the obtained dispersion relation in the inertial range is nonlinear, though it is only weakly nonlinear. It shows the increase of wavenumber with frequency, i.e. the decrease in scales with increasing frequency which is in good approximate agreement with Kolmogorov's prediction. From the dispersion relation we find that the shorter scales disperse at higher phase and group velocities than the longer scales. This kind of dispersion is typical for progressively steepened gradients and formation of narrow structures. Hence the dispersion relation obtained from the observations supports the conclusion suggested by the increase in conductivity with frequency that the inertial range is the range where the turbulence generates structures on the narrow scales. 

A further practical conclusion drawn from the rather weak nonlinear dependence of frequency from wavenumber is the (completely independent) observation-based \emph{confirmation} of Taylor's hypothesis which implies that in a plasma streaming with speed $|\mathbf{v}|\gg v_\mathit{ph}$ one can put $\omega -kv_\mathrm{ph}=\mathbf{k\cdot v}$ in order to transform from frequency to wavenumber spectra. Use of the obtained dependence $\omega(k)$ should produce a more precise spectral shape when applying Taylor's hypothesis. 

It should, however, be stressed that the observations used do not distinguish between the mean-field parallel and perpendicular components. This means that our results are a mixture of the effects of the turbulent fluctuations in all directions. They are thus to some sense average. They require observational improvement and separation of the compressional magnetic component. Its interpretation requires a different treatment. 

It would be interesting to extend investigations of this kind to other parts of the turbulent spectrum, the range between energy injection and the possibly different inertial ranges as well as extension into the dissipation range. This requires knowledge of the magnetic and also the electric power spectra. Observations of electric power spectra in the dissipation range will, however, be difficult to perform in the solar wind. They require very high (on the electron gyration timescale) resolution of the electron and ion distribution functions in order to be able to perform the transformation of the electric field into the frame of the solar wind. Direct observations of the electric mean and fluctuation fields using injection of gyrating ion beams would be a more promising way. Some thought in this direction, particularly the dissipation range, will become included elsewhere \citep{treumann2017}.

\begin{acknowledgement}
This research was part of a Visiting Senior Scientist Programme at ISSI, Bern. We recognize the interest of the ISSI directorate. Hospitality of the friendly ISSI staff is thankfully acknowledged.
\end{acknowledgement}

\end{document}